\documentclass[%
aps,
prl,%
amssymb,
reprint,%
citeautoscript,
]{revtex4-1}

\usepackage{color}
\usepackage{graphicx}% Include figure files
\usepackage{dcolumn}% Align table columns on decimal point
\usepackage{bm}% bold math
\usepackage{amssymb}
\usepackage{amsmath}

\usepackage{eso-pic}
\usepackage{fix-cm}

\usepackage{floatrow}
\usepackage{marginnote}
\usepackage{soul}

\newcommand{\sss}{\scriptscriptstyle}
\newcommand{\sst}{\scriptstyle}

\newcommand{\stext}[1]{\sss \text{#1} \sst}

\begin{document}
\graphicspath{{fig/}}

\title{Polarization selection rules for inter-Landau level transitions in epitaxial graphene revealed by infrared optical Hall effect}

\author{P.~K\"{u}hne}
   \email{kuehne@huskers.unl.edu}
   \homepage{http://ellipsometry.unl.edu}
   \affiliation{Department of Electrical Engineering and Center for Nanohybrid Functional Materials, University of Nebraska-Lincoln, U.S.A.}
\author{V.~Darakchieva}
   \affiliation{\mbox{Department of Physics, Chemistry and Biology, IFM, Link\"{o}ping University, SE-581 83 Link\"{o}ping, Sweden}}
\author{J.D.~Tedesco}
   \affiliation{ABB, Inc.~171 Industry Drive, P.O. Box 38, Bland, VA 24315, U.S.A.}
\author{R.L.~Myers-Ward}
   \affiliation{U.S.~Naval Research Laboratory, Washington, DC 20375, U.S.A.}
\author{C.R.~Eddy~Jr.}
   \affiliation{U.S.~Naval Research Laboratory, Washington, DC 20375, U.S.A.}
\author{D.K.~Gaskill}
   \affiliation{U.S.~Naval Research Laboratory, Washington, DC 20375, U.S.A.}
\author{R.~Yakimova}
   \affiliation{\mbox{Department of Physics, Chemistry and Biology, IFM, Link\"{o}ping University, SE-581 83 Link\"{o}ping, Sweden}}
\author{C.M.~Herzinger}
   \affiliation{\mbox{J.A.~Woollam Co.,~Inc.,~645 M Street, Suite 102, Lincoln, NE 68508-2243, U.S.A.}}
\author{J.A.~Woollam}
   \affiliation{\mbox{J.A.~Woollam Co.,~Inc.,~645 M Street, Suite 102, Lincoln, NE 68508-2243, U.S.A.}}
\author{M.~Schubert}
   \affiliation{Department of Electrical Engineering and Center for Nanohybrid Functional Materials, University of Nebraska-Lincoln, U.S.A.}
\author{T.~Hofmann}
   \affiliation{Department of Electrical Engineering and Center for Nanohybrid Functional Materials, University of Nebraska-Lincoln, U.S.A.}

\begin{abstract}
We report on polarization selection rules of inter-Landau level transitions using reflection-type optical Hall effect measurements from 600 to 4000~cm$^{-1}$ on epitaxial graphene grown by thermal decomposition of silicon carbide. We observe symmetric and anti-symmetric signatures in our data due to polarization preserving and polarization mixing inter-Landau level transitions, respectively. From field-dependent measurements we identify that transitions in coupled graphene mono-layers are governed by polarization mixing selection rules, whereas transitions in decoupled graphene mono-layers are governed by polarization preserving selection rules. The selection rules may find explanation by different coupling mechanisms of inter-Landau level transitions with free charge carrier magneto-optic plasma oscillations.

\end{abstract}


\maketitle

Epitaxial graphene grown by thermal decomposition onto SiC substrates has received tremendous interest due to its unique physical and electronic properties where free charge carriers behave as quasi Dirac particles, for instance~\cite{VirojanadaraPRB78_2008, LinIEDL31_2010, deHeerPNAS108_2011, LinS332_2011, WuNL12_2012,MorimotoPRB86_2012,HenriksenPRL100_2008, CrasseeNP7_2011,CrasseePRB84_2011,TzalenchukNN5_2010, BergerJPCB108_2004, BergerS312_2006}. Infrared magneto-optic spectroscopy has been widely applied to probe the electronic states of graphene by monitoring the magnetic field and frequency dependencies of inter-Landau level transitions~\cite{BergerJPCB108_2004, BergerS312_2006,SadowskiPRL97_2006,OrlitaPRL101_2008,OrlitaPRL107_2011}. However, the polarization properties of inter-Landau level transitions, and their polarization selection rules, that is, whether individual transitions are polarization preserving or polarization mixing, are unknown. The polarization selection rules determine the symmetry properties of the magneto-optic dielectric permittivity tensor $\bm{\varepsilon^{\stext{\bf{MO}}}} $. For a given series of Landau level transitions, $\bm{\varepsilon^{\stext{\bf{MO}}}} $ can be constructed, and compared with physical model descriptions. Sufficient information for obtaining the polarization selection rules, and to construct $\bm{\varepsilon^{\stext{\bf{MO}}}} $, is provided by optical Hall effect measurements~\cite{HofmannRSI77_2006,HofmannTSF519_2011}, and is presented in this letter. We report here on our observation of isotropic and anisotropic Landau transitions from optical Hall effect measurements which differ in their magnetic field dependencies, and we discuss possible physical origins by reconstructing $\bm{\varepsilon^{\stext{\bf{MO}}}} $ using simple model scenarios.

\begin{figure*}
\floatbox[{\capbeside\thisfloatsetup{capbesideposition={right,center},capbesidewidth=4.5cm}}]{figure}[\FBwidth]
{\caption{Optical Hall effect experimental data (green) and best-model fit data (red) for epitaxial graphene at $B_c$=+(5.66$\pm$0.02)~T and $T=1.5$~K. The angle of incidence is $\varPhi_{\stext{a}}=45^{\circ}$. DTC, BLG and SLG denote contributions assigned in this letter to Drude-type carriers (DTC), bi-layer graphene LL (BLG) and single-layer graphene LL (SLG), respectively. Signatures indicated by SLG are isotropic and do not occur in off-diagonal-block elements $\delta \mathrm{M_{13}}$, $\delta \mathrm{M_{23}}$, $\delta \mathrm{M_{31}}$, and $\delta \mathrm{M_{32}}$ (shaded background), while features labeled with DTC and BLG are anisotropic and cause polarization mixing.}
\label{fig:MM_8T}}
{\includegraphics[keepaspectratio=true,trim=0 0 0 0, clip,width=11cm]{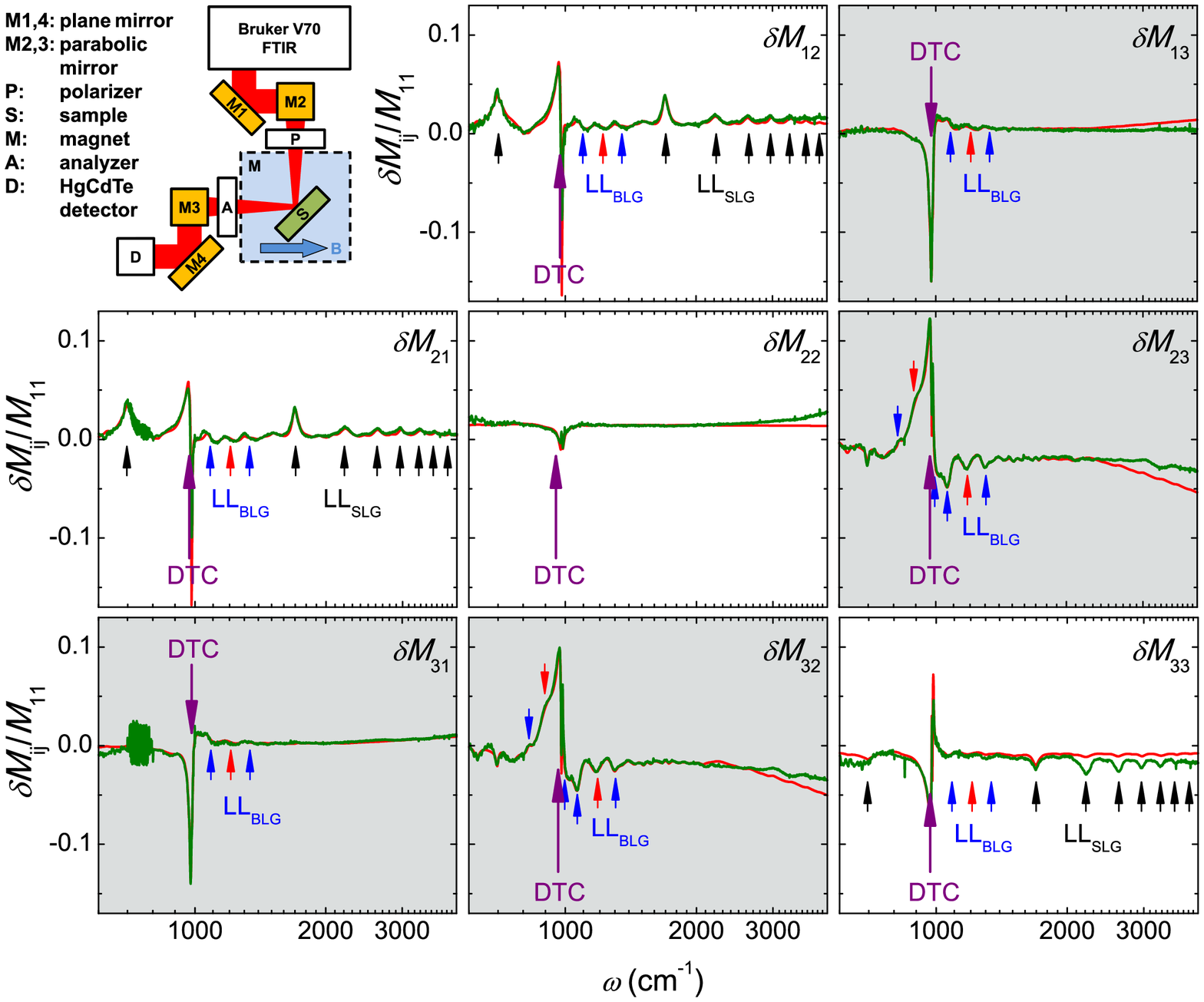}}\vspace{-0.2cm}
\end{figure*}

The optical Hall effect determines changes of optical properties of thin film samples under the influence of external magnetic fields~\cite{SchubertJOSAA20_2003,Hofmannpss205_2008,HofmannTSF519_2011}. In contrast to Faraday rotation, measurements are conveniently taken in reflection-type arrangement at oblique angle of incidence, thereby discriminating between parallel and perpendicular polarization. Measurements are performed in the Stokes vector approach, and results are reported in the Mueller matrix presentation, which allows immediate differentiation between polarization preserving (isotropic) as well as polarization mixing (anisotropic) sample properties. Subsequent data analysis using model approaches for the dielectric function, or equivalently the optical conductivity, permits quantitative access to physical model parameters. In the Stokes formalism the Mueller matrix $\mathbf{M}$ connects the Stokes vector of incident and reflected electromagnetic waves $\mathbf{S}^{\stext{in}}$ and $\mathbf{S}^{\stext{out}}$, respectively, where $\mathbf{S}^{\stext{out}}=\mathbf{M}\mathbf{S}^{\stext{in}}$~\cite{Fujiwara_2007}. The optical Hall effect determines magnetic field-induced differences $\delta \mathbf{M}$ with respect to $\mathbf{M}$ at zero field. $\delta \mathbf{M}$ contains non-zero on-diagonal block elements only ($\delta M_{\mathrm{11}}$, $\delta M_{\mathrm{12}}$, $\delta M_{\mathrm{21}}$, $\delta M_{\mathrm{22}}$, $\delta M_{\mathrm{33}}$, $\delta M_{\mathrm{34}}$, $\delta M_{\mathrm{43}}$, $\delta M_{\mathrm{44}}$) when the magnetic field induced sample response is purely isotropic, i.e. no polarization mixing occurs. Additional, non-zero off-diagonal block elements occur ($\delta M_{\mathrm{13}}$, $\delta M_{\mathrm{31}}$, $\delta M_{\mathrm{14}}$, $\delta M_{\mathrm{41}}$, $\delta M_{\mathrm{23}}$, $\delta M_{\mathrm{32}}$, $\delta M_{\mathrm{24}}$, $\delta M_{\mathrm{42}}$) when the magnetic field induced sample response is anisotropic, i.e., polarization mode mixing occurs~\cite{Fujiwara_2007}. Therefore, symmetric signatures of Landau level transitions occur within the on-diagonal block elements only. Anti-symmetric signatures occur also within all off-diagonal-block elements. We note that in our setup elements of both 4$^{th}$ row and column are inaccessible, and all remaining elements are normalized by $M_{\mathrm{11}}$ removing light source base line fluctuations, providing 8 independent pieces of information.

The epitaxial graphene sample investigated in this letter was grown by sublimation on the C-polar (000$\bar{1}$) surface of a semi-insulating 6H-SiC substrate. During the growth, the SiC substrate was heated to 1400~$^\circ$C in argon atmosphere. Further information on growth conditions can be found in Ref.~\onlinecite{TedescoAPL96_2010}. We estimate the number of graphene layers to be 10-20, similar to those measured previously on C-face 4H-SiC~\cite{BoosalisAPL101_2012}. Optical Hall effect measurements where carried out at $\mathit{\Phi}_a$~=~45$^\circ$ angle of incidence in the spectral range from 600 to 4000~cm$^{-1}$ with spectral resolution of 1~cm$^{-1}$~\cite{HofmannRSI77_2006}. The sample was held at temperature $T=1.5$~K. The magnetic field was varied from $B=0$~T to 8~T in 0.1~T increments while the magnetic field direction was parallel to the reflected beam resulting in a magnetic field $B_c=B$/$\sqrt{2}$ parallel to the sample normal.

Quantitative optical Hall effect data analysis requires stratified layer model calculations in which parametrized dielectric functions are used. Least-square principle methods are employed in order to vary model parameters until calculated and experimental ellipsometric data are matched as closely as possible (best-model)~\cite{SchubertIRSEBook_2004}. The dielectric functions for silicon carbide are composed of Lorentzian-broadened oscillators described by the respective longitudinal-optical (LO) and transverse-optical (TO) phonon frequencies~\cite{TiwaldPRB60_1999}.

Figure~\ref{fig:MM_8T} depicts results of optical Hall effect measurements at $B_c = +(5.66 \pm 0.02)$~T. Graphs are arranged according to their appearance in the Mueller matrix, with the top left corner ($\delta M_{\mathrm{11}}$) omitted. The spectral response observed in the on-diagonal block elements is distinctly different from the off-diagonal block response. Multiple transitions with different polarization signatures can be identified. Sets of signatures belong to different series of Landau level transitions, as will be discussed below. Comparing representative elements, e.g., $\delta M_{33}$ and $\delta M_{32}$, two sets with different polarization properties can be identified. The first set of transitions, indicated with vertical arrows labeled LL$_{\stext{SLG}}$, is isotropic, i.e., not associated with polarization mixing, and spread out over the entire measured spectral range. These transitions do not occur in off-diagonal block elements. The term ``SLG'' stands for single-layer graphene, as discussed further below.  The second set of resonances, indicated with vertical arrows labeled LL$_{\stext{BLG}}$, is anisotropic, i.e., associated with polarization mixing, and occurs in a narrower range from 600 to 1500~cm$^{-1}$. The term ``BLG'' indicates bi-layer graphene as discussed further below as well. However, a subset of these transitions will be assigned to tri-layer graphene. In addition, a pronounced feature observed at 970~cm$^{-1}$, indicated with vertical arrows and labeled by DTC, is common to all graphs in Fig.~\ref{fig:MM_8T} and hence anisotropic. The term ``DTC'' stands for Drude-type carriers. Non-zero off-diagonal block Mueller matrix elements are inherently tied to the existence of off-diagonal components in $\bm{\varepsilon^{\stext{\bf{MO}}}} $. Thus, a priori, one must conclude that transitions LL$_{\stext{SLG}}$ originate from processes in sample regions with polarizability contributions to $\bm{\varepsilon^{\stext{\bf{MO}}}} $ that are diagonal in $\bm{\varepsilon^{\stext{\bf{MO}}}} $ and hence isotropic. Likewise, transitions LL$_{\stext{BLG}}$ are to be described by contributions with non-diagonal components in $\bm{\varepsilon^{\stext{\bf{MO}}}} $.

\begin{figure}
	\caption{Selected on- and off-diagonal-block optical Hall effect spectra ($T$~=~1.5~K, $\mathit{\Phi}_a$~=~45$^\circ$) for $B_c~=+(0.707\pm 0.002)$~T to $B_c~=+(5.66\pm 0.02)$~T in 0.07~T increments. The graphs are stacked by 0.006. (a) $\delta M_{33}$; Isotropic Landau level transitions, indicated by letters according to Sadowski \textit{et al.}~\cite{SadowskiPRL97_2006} (b) $\delta M_{32}$; Anisotropic Landau transitions, indicated by blue and red arrows. \vspace{-.7cm}}
	\label{fig:MM33_32}
	\centering
	\includegraphics[keepaspectratio=true,trim=0 0 0 0, clip,width=8.5cm]{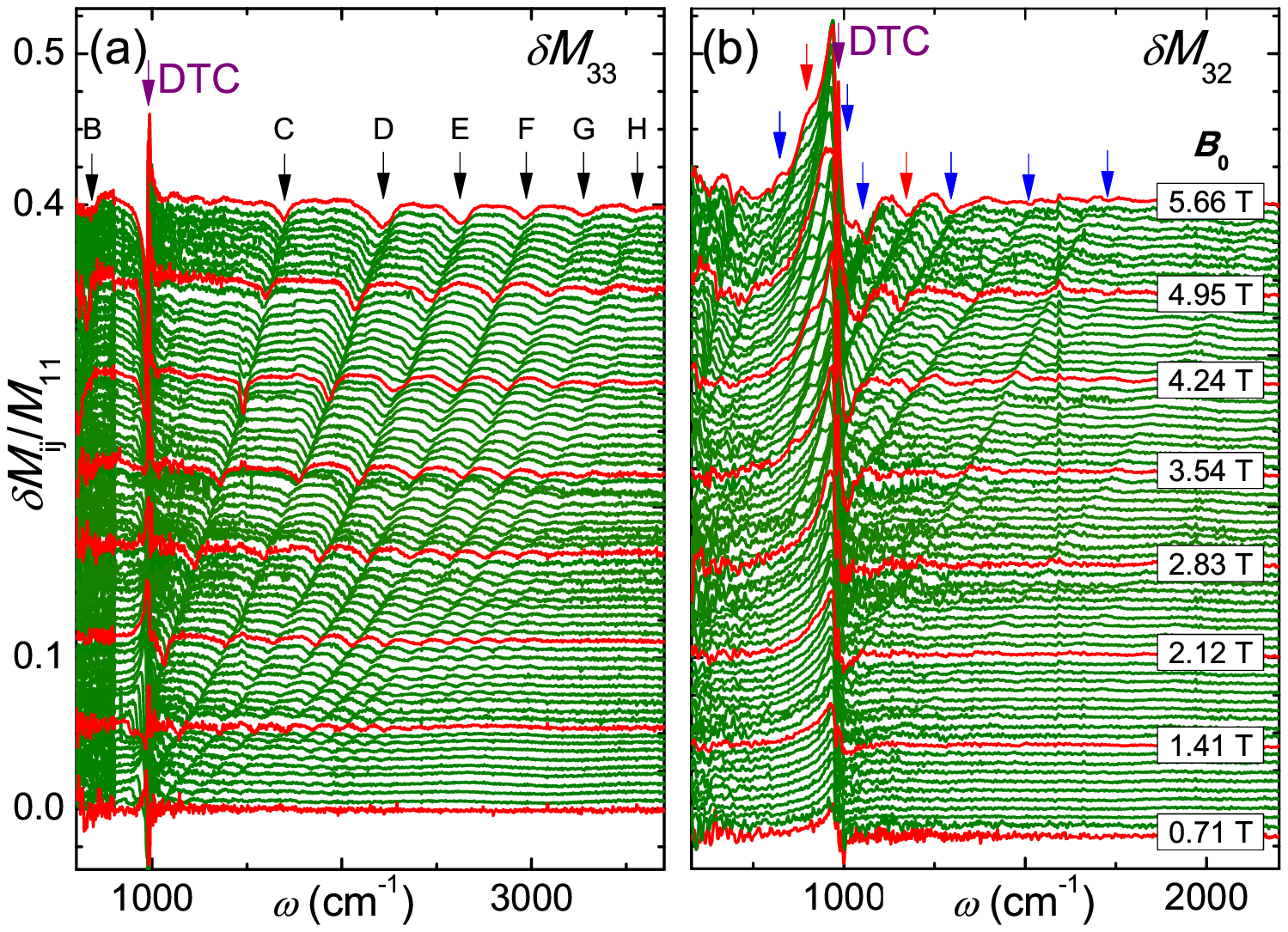}
	\vspace{-0.3cm}
\end{figure}

Without loss of generality, the optical response of electronic systems with bound and unbound excitations subjected to external magnetic fields can be constructed by using magneto-optic polarizability functions $\chi_{\mathrm{+}}$ and $\chi_{\mathrm{-}}$ for right- and left-handed circularly polarized light, respectively~\cite{SchubertJOSAA20_2003}. In Cartesian coordinates $\bm{\varepsilon^{\stext{\bf{MO}}}} $ is then anti-symmetric:

\begin{equation} \label{eq:LLTeps}
	\bm{\varepsilon^{\stext{\bf{MO}}}}
	=
	\textbf{I}+\frac{1}{2}
	\begin{pmatrix}
		\;\;\;\;(\chi_{\mathrm{+}}+\chi_{\mathrm{-}})	& \mathrm{i}(\chi_{\mathrm{+}}-\chi_{\mathrm{-}}) & 0 \\ 			 -\mathrm{i}(\chi_{\mathrm{+}}-\chi_{\mathrm{-}}) &	 \:(\chi_{\mathrm{+}}+\chi_{\mathrm{-}}) & 0 \\
		0	& 0 & 0 \\
	\end{pmatrix}\,,
\end{equation}
where \textbf{I} indicates the unit matrix and the magnetic field is taken along the $z$-direction. All other dielectric contributions are omitted. For $\chi_{\mathrm{+}} \neq \chi_{\mathrm{-}}$, $\bm{\varepsilon^{\stext{\bf{MO}}}}$ describes a medium with anisotropic magneto-optical properties which produce non-vanishing off-diagonal block elements in the optical Hall effect, whereas for $\chi_{\mathrm{+}} = \chi_{\mathrm{-}}$, $\bm{\varepsilon^{\stext{\bf{MO}}}}$ describes a medium with isotropic magneto-optical properties, and off-diagonal block elements in the optical Hall effect do not occur. Hence, for transitions LL$_{\stext{SLG}}$ $\chi_{\mathrm{+}} = \chi_{\mathrm{-}}$, while for LL$_{\stext{BLG}}$ $\chi_{\mathrm{+}} \neq \chi_{\mathrm{-}}$. A semi-classical description for $\chi_{\mathrm{\pm}}$ can be obtained using $n$ series of Lorentzian-broadened Green functions at energies $\hbar\omega_{0,n}$ with spectral weight $\omega^2_{p,n}$, scattering life time $1/\gamma_{n}$ and including coupling to a magneto-optic plasma mode $\omega_{c}$
\begin{equation} \label{eq:chi-model}
	\chi_{\mathrm{\pm}}=\sum_n\omega^2_{p,n}\left(\omega^2_{0,n}-\omega^2-i\omega \gamma_{p,n} \pm i\omega \omega_{c}\right)^{-1}.
\end{equation}

When the plasma coupling is turned off, i.e., $\omega_{c}\rightarrow0$, $\bm{\varepsilon^{\stext{\bf{MO}}}} $ becomes symmetric. Standard layer model calculations to determine the Mueller matrix elements of epitaxial graphene on silicon carbide using Eqs.~(\ref{eq:LLTeps}) and (\ref{eq:chi-model}) reproduce line shape and isotropy of all features labeled LL$_{\stext{SLG}}$ in Fig.~\ref{fig:MM_8T}. When coupling with plasma modes is considered, i.e., $\omega_{c} > 0$, features are mapped out onto the off-diagonal-block elements, and line shapes match excellently with the experimental data for transitions labeled by LL$_{\stext{BLG}}$.

Figure~\ref{fig:MM33_32} summarizes representative on-diagonal and off-diagonal-block Mueller matrix difference spectra as a function of the magnetic field. The anisotropic resonance at 970~cm$^{-1}$, labeled with DTC, increases in amplitude with increasing magnetic field strength. The wavenumber, however, at which this resonance occurs does not vary with the external magnetic field strength. The physical origin of this resonance is the coupled motion of bound charge displacement near the longitudinal optical phonon mode of the silicon carbide substrate with a free charge carrier plasma at the sample surface producing resonant magneto-optic birefringence~\cite{SchubertIRSEBook_2004, SchubertPRB71_2005, SchubertJOSAA20_2003}. Eqs.~(\ref{eq:LLTeps}) and (\ref{eq:chi-model}) can be used to render this phenomenon when $\hbar\omega_{0}$ is the (bound) phonon mode energy. The resulting polarization contribution is anisotropic and occurs in all Mueller matrix elements. We attribute this mode to highly doped graphene layers in close vicinity of the interface between the substrate and the epitaxial graphene, originating from SiC charge transfer~\cite{LinAPL97_2010}.

\begin{figure}
 \includegraphics[keepaspectratio=true,trim=0 -10 0 0, clip,width=8.5cm]{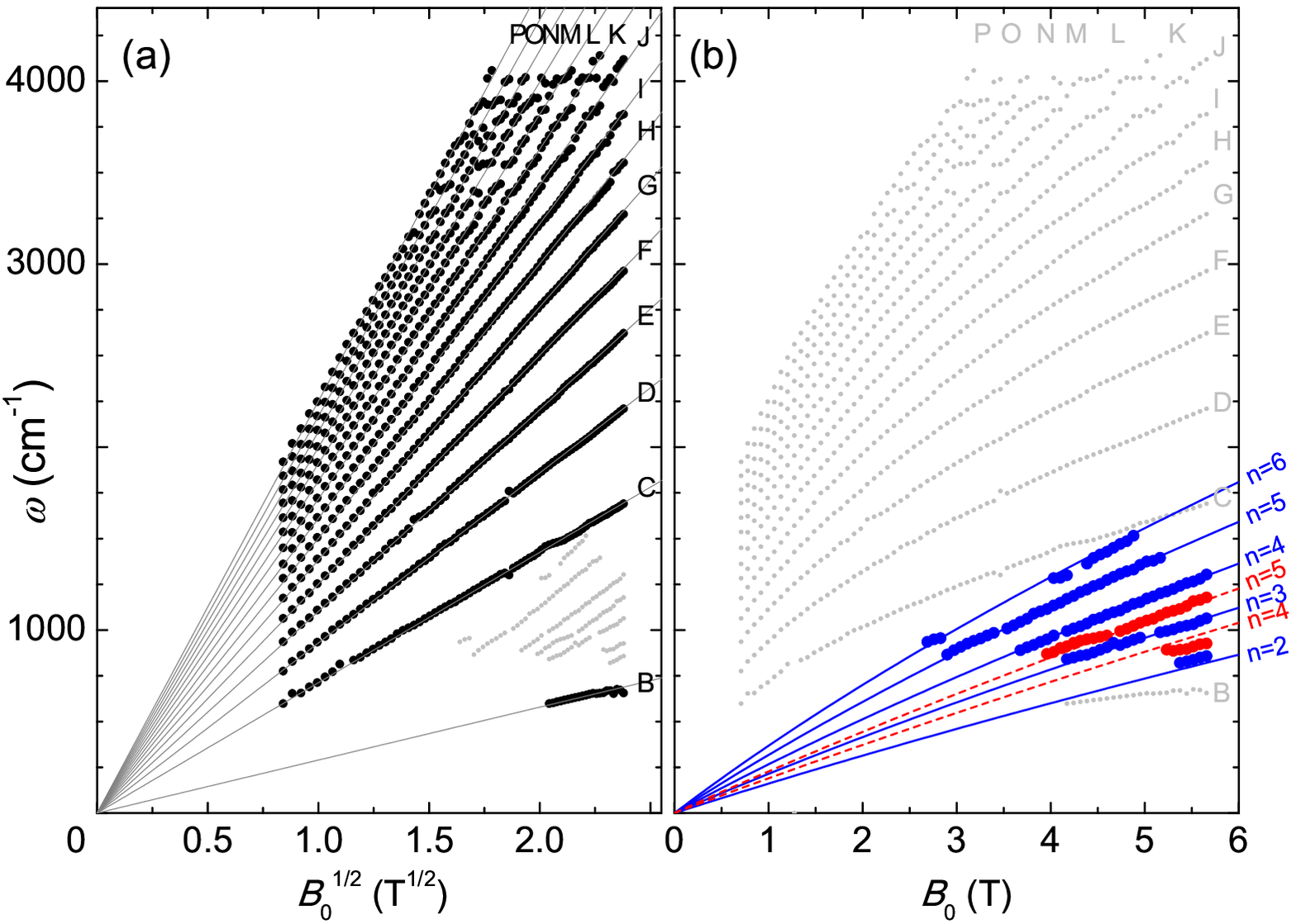}
 \vspace{-0.5cm}
	\caption{Symmetric (a; isotropic) and anti-symmetric (b; anisotropic) Landau transition energies in epitaxial graphene determined from optical Hall effect measurements at 1.5~K. Solid lines denote single-layer graphene (a; black), bi-layer (b; blue) and tri-layer graphene (b; red) dependencies on the magnetic field. \vspace{-.7cm}}
 \label{fig:results}
 \centering
\end{figure}

The magnetic field-dependent measurements reveal that energy spacings of transitions LL$_{\stext{SLG}}$ scale with the square root of transition index $n$ and magnetic field $B$, indicative for the Dirac-type bandstructure with Fermi level close to the charge neutrality point~\cite{SadowskiPRL97_2006}. Data in Fig.~\ref{fig:results}~(a) are parameters $\omega_{0,n}$ obtained from best-match model analysis of the optical Hall effect spectra as a function of $B_c$. We attribute these transition to originate within regions of the epitaxial graphene that are composed of decoupled, quasi neutral graphene sheets~\cite{OrlitaPRL101_2008}. Energies in Fig.~\ref{fig:results}~(a) follow $E_{\stext{SLG}}^{\stext{LL}}(n) =\mbox{sign}(n)E_0\sqrt{|n|}$ with $E_0=\tilde{c}\sqrt{2\hbar e|B_c|}$ and average velocity of Dirac fermions $\tilde{c}$. The naming convention used to indicate transitions in Fig.~\ref{fig:results}(a) is adapted from Sadowski \it et~al\rm~\cite{SadowskiSSC143_2007}. Optical selection rules for transitions between levels $n'$ and $n$ require $|n'|=|n|\pm 1$. The best-match model velocity obtained from matching all data in Fig.~\ref{fig:results}~(a) is $\tilde{c}=(1.01\pm0.01)\times 10^6$m/s, in very good agreement with Refs.~\onlinecite{OrlitaPRB83_2011, SadowskiPRL97_2006, OrlitaPRL102_2009, HenriksenPRL100_2008, OrlitaPRL107_2011, OrlitaPRL101_2008}. The corresponding best-match functions $E_{\stext{SLG}}^{\stext{LL}}$ versus $B_c$ are plotted as solid lines in Fig.~\ref{fig:results} (a).

Transition energy parameters obtained from best-match model analysis of the off-diagonal-block optical Hall effect data are plotted in Fig.~\ref{fig:results}~(b) and exhibit sublinear behavior. The magnetic field scaling of the energy spacings for the anisotropic transitions suggests bi- and tri-layer graphene as their physical origin. The transition energies of N-layer graphene have been described as~\cite{AbergelPRB75_2007,KoshinoPRB77_2008, OrlitaPRL107_2011}
\vspace{-0.2cm}
\begin{equation} \label{eqn:LLNLayer}
\begin{split}
	E^{\stext{LL}}_{\stext{N-BLG}}&(n,\mu)
	= \mbox{sign}(n)\frac{1}{\sqrt{2}}\left[\vphantom{\frac{1}{\sqrt{2}}}(\lambda_{\stext{N}}\gamma)^2+\left(2\left| n\right|+1\right)E_0^2\right. \\
	& \hspace{-6mm} \left. +\mu\sqrt{(\lambda_{\stext{N}}\gamma)^4+2\left(2\left| n\right|+1\right)E_0^2(\lambda_{\stext{N}}\gamma)^2+E_0^4}\right]^{1/2},
\end{split}
\end{equation}
\noindent with coupling constant $\gamma$, layer parameter $\lambda_{\stext{N}}$~\cite{KoshinoPRB77_2008} and where $\mu=-1,+1$ corresponds to the higher and lower sub-bands in the limit of zero magnetic field, respectively~\cite{KoshinoPRB77_2008}. Optical selection rules are the same as for $E_{\stext{SLG}}^{\stext{LL}}(n)$. Using $\tilde{c}=(1.01\pm0.01)\times 10^6$m/s, best-match model functions are plotted in Fig.~\ref{fig:results}(b) for $N=2$ and $N=3$ as blue solid and red dashed lines, respectively. No transition was observed that would correspond to $N>3$. These transitions can be assigned to Bernal stacked bi-layer graphene ($N=2$) and tri-layer graphene ($N=3$). The best-match model parameters obtained here are $\gamma=(3120\pm 175)$~cm$^{-1}$ for $N=2$, corroborating the result obtained by Orlita \it et al.\rm\ from FTIR transmission experiments~\cite{OrlitaPRL107_2011}. For $N=3$, we observe $\gamma=(3150\pm 20)$~cm$^{-1}$, which render the first experimental confirmation of the theoretical predictions by Koshino and Ando for tri-layer graphene~\cite{KoshinoPRB77_2008}. We note that these anisotropic Landau transitions are only observed and resolved for $n=2\ldots 6$ for bi-layer graphene, and for $n=4$ and $n=5$  for tri-layer graphene. At this point, we do not know why transitions with  $n>6$ for $N=2$ and $n \neq 4, n \neq 5$ for $N=3$ cannot be observed. However, the fact that these transitions appear with anisotropic optical Hall effect signatures suggests their coupling with free charge carriers within the sample. We propose that stacking order defects within the bi-layer and tri-layer graphene allow for coupling of Dirac particles within their Landau levels with the cyclotron resonance~\cite{CrasseeNP7_2011} of the free carrier plasma resulting in anisotropic Landau transition signatures.

The polarization selection rules are obtained here from polarization resolved measurements of the Optical Hall effect instrument. The different polarization selection rules observed for the transitions labeled with SLG and BLG indicate that different physical processes contribute to their respective magneto-optic polarizability tensor. A possible mechanism using coupled or uncoupled bound and unbound electronic transitions was described above. According to this scenario transitions BLG are affected by free charge carrier plasma oscillations whereas transitions SLG are not. The different coupling mechanisms for transitions in sets SLG and BLG can be concluded here from the knowledge of the polarization selection rules. We consider it worth noting that the polarization selection rules observed here for the different sets of Landau transitions are invariant with respect to the magnetic field strength, over the range of magnetic fields observed here. This suggests that the mechanisms which lead to the different polarization selection rules, for example the coupling of bound electronic transitions with magnetoplasma oscillation is not affected by the  magnetic field, and which may, however, change at larger magnetic fields than those investigated here.

In conclusion, we find that inter-Landau level transitions in epitaxial graphene grown on C-face SiC are governed by different polarization selection rules. The transitions belong to different sets, where each set possesses their own polarization selection rule. Hence, for a given magnetic field, the polarization behavior and selection rules can be used to differentiate the sets to which observed transitions belong. The polarization behavior can be obtained, for example, by reflection-type optical Hall effect measurements in the infrared spectral region. Specifically, for epitaxial graphene, two sets of transition series with polarization preserving as well as polarization mixing properties occur. We identify that the polarization preserving transitions originate from decoupled graphene mono-layers, while the polarization mixing transitions originate from bi-layer and tri-layer graphene. This identification follows from observation of transitions belonging to equal polarization rules as a function of the magnetic field strengths. The polarization preserving and polarization mixing rules may find explanation by different coupling mechanisms of inter-Landau level transitions with free charge carrier magneto-optic plasma oscillations.

The authors would like to acknowledge financial support from the Army Research Office (D. Woolard, Contract No. W911NF-09-C-0097), the National Science Foundation (Grant Nos. MRSEC DMR-0820521, MRI DMR-0922937, EPS-1004094, and DMR-0907475), the University of Nebraska-Lincoln, the J.A. Woollam Foundation, the Office of Naval Research, the Swedish Research Council (VR) under grant No.2010-3848 and, the Swedish Governmental Agency for Innovation Systems (VINNOVA) under the VINNMER international qualification program, grant No.2011-03486. \vspace{-.5cm}

\vspace{-0.1cm}

%\bibliographystyle{APLstyleNOwebref}
%\bibliography{CompleteLibrary}

\begin{thebibliography}{10}
\providecommand{\bibnamefont}[1]{#1}
\providecommand{\bibfnamefont}[1]{#1}

\bibitem{BergerJPCB108_2004}
\bibfnamefont{C.}~\bibnamefont{Berger}, \bibfnamefont{Z.}~\bibnamefont{Song},
  \bibfnamefont{T.}~\bibnamefont{Li}, \bibfnamefont{X.}~\bibnamefont{Li},
  \bibfnamefont{A.~Y.} \bibnamefont{Ogbazghi},
  \bibfnamefont{R.}~\bibnamefont{Feng}, \bibfnamefont{Z.}~\bibnamefont{Dai},
  \bibfnamefont{A.~N.} \bibnamefont{Marchenkov}, \bibfnamefont{E.~H.}
  \bibnamefont{Conrad}, \bibfnamefont{P.~N.} \bibnamefont{First},
  \bibnamefont{and} \bibfnamefont{W.~A.} \bibnamefont{de~Heer}, J. Phys. Chem.
  B \textbf{108}, 19912 (2004).

\bibitem{BergerS312_2006}
\bibfnamefont{C.}~\bibnamefont{Berger}, \bibfnamefont{Z.}~\bibnamefont{Song},
  \bibfnamefont{X.}~\bibnamefont{Li}, \bibfnamefont{X.}~\bibnamefont{Wu},
  \bibfnamefont{N.}~\bibnamefont{Brown}, \bibfnamefont{C.}~\bibnamefont{Naud},
  \bibfnamefont{D.}~\bibnamefont{Mayou}, \bibfnamefont{T.}~\bibnamefont{Li},
  \bibfnamefont{J.}~\bibnamefont{Hass}, \bibfnamefont{A.~N.}
  \bibnamefont{Marchenkov}, \bibfnamefont{E.~H.} \bibnamefont{Conrad},
  \bibfnamefont{P.~N.} \bibnamefont{First}, \emph{et~al.}, Science
  \textbf{312}, 1191 (2006).

\bibitem{VirojanadaraPRB78_2008}
\bibfnamefont{C.}~\bibnamefont{Virojanadara},
  \bibfnamefont{M.}~\bibnamefont{Syvaejarvi},
  \bibfnamefont{R.}~\bibnamefont{Yakimova}, \bibfnamefont{L.~I.}
  \bibnamefont{Johansson}, \bibfnamefont{A.~A.} \bibnamefont{Zakharov},
  \bibnamefont{and} \bibfnamefont{T.}~\bibnamefont{Balasubramanian}, Phys. Rev.
  B \textbf{{78}}, 245403 ({2008}).

\bibitem{HenriksenPRL100_2008}
\bibfnamefont{E.~A.} \bibnamefont{Henriksen},
  \bibfnamefont{Z.}~\bibnamefont{Jiang}, \bibfnamefont{L.-C.}
  \bibnamefont{Tung}, \bibfnamefont{M.~E.} \bibnamefont{Schwartz},
  \bibfnamefont{M.}~\bibnamefont{Takita}, \bibfnamefont{Y.-J.}
  \bibnamefont{Wang}, \bibfnamefont{P.}~\bibnamefont{Kim}, \bibnamefont{and}
  \bibfnamefont{H.~L.} \bibnamefont{Stormer}, Phys. Rev. Lett. \textbf{100},
  087403 (2008).

\bibitem{TzalenchukNN5_2010}
\bibfnamefont{A.}~\bibnamefont{Tzalenchuk},
  \bibfnamefont{S.}~\bibnamefont{{Lara-Avila}},
  \bibfnamefont{A.}~\bibnamefont{Kalaboukhov},
  \bibfnamefont{S.}~\bibnamefont{Paolillo},
  \bibfnamefont{M.}~\bibnamefont{Syvajarvi},
  \bibfnamefont{R.}~\bibnamefont{Yakimova},
  \bibfnamefont{O.}~\bibnamefont{Kazakova}, \bibfnamefont{J.~J.~B.}
  \bibnamefont{M.}, \bibfnamefont{V.}~\bibnamefont{Fal'ko}, \bibnamefont{and}
  \bibfnamefont{S.}~\bibnamefont{Kubatkin}, Nat Nano \textbf{5}, 186 (2010).

\bibitem{LinIEDL31_2010}
\bibfnamefont{Y.-M.} \bibnamefont{Lin}, \bibfnamefont{H.-Y.}
  \bibnamefont{Chiu}, \bibfnamefont{K.~A.} \bibnamefont{Jenkins},
  \bibfnamefont{D.~B.} \bibnamefont{Farmer},
  \bibfnamefont{P.}~\bibnamefont{Avouris}, \bibnamefont{and}
  \bibfnamefont{A.}~\bibnamefont{Valdes-Garcia}, IEEE Electron. Device Letters
  \textbf{31}, 68 (2010).

\bibitem{deHeerPNAS108_2011}
\bibfnamefont{W.}~\bibnamefont{de~Heer},
  \bibfnamefont{C.}~\bibnamefont{Berger}, \bibfnamefont{M.}~\bibnamefont{Ruan},
  \bibfnamefont{M.}~\bibnamefont{Sprinkle}, \bibfnamefont{X.}~\bibnamefont{Li},
  \bibfnamefont{Y.}~\bibnamefont{Hu}, \bibfnamefont{B.}~\bibnamefont{Zhang},
  \bibfnamefont{J.}~\bibnamefont{Hankinson}, \bibnamefont{and}
  \bibfnamefont{E.}~\bibnamefont{Conrad}, PNAS \textbf{108}, 16900 (2011).

\bibitem{LinS332_2011}
\bibfnamefont{Y.~M.} \bibnamefont{Lin},
  \bibfnamefont{A.}~\bibnamefont{Valdes-Garcia}, \bibfnamefont{S.~J.}
  \bibnamefont{Han}, \bibfnamefont{D.~B.} \bibnamefont{Farmer},
  \bibfnamefont{I.}~\bibnamefont{Meric}, \bibfnamefont{Y.~N.}
  \bibnamefont{Sun}, \bibfnamefont{Y.~Q.} \bibnamefont{Wu},
  \bibfnamefont{C.}~\bibnamefont{Dimitrakopoulos},
  \bibfnamefont{A.}~\bibnamefont{Grill}, \bibnamefont{and} \bibfnamefont{P.~A.
  K.~A.} \bibnamefont{Jenkins}, Science \textbf{332}, 6035 (2011).

\bibitem{CrasseeNP7_2011}
\bibfnamefont{I.}~\bibnamefont{Crassee},
  \bibfnamefont{J.}~\bibnamefont{Levallois}, \bibfnamefont{A.~L.}
  \bibnamefont{Walter}, \bibfnamefont{M.}~\bibnamefont{Ostler},
  \bibfnamefont{A.}~\bibnamefont{Bostwick},
  \bibfnamefont{E.}~\bibnamefont{Rotenberg},
  \bibfnamefont{T.}~\bibnamefont{Seyller},
  \bibfnamefont{D.}~\bibnamefont{van~der Marel}, \bibnamefont{and}
  \bibfnamefont{A.~B.} \bibnamefont{Kuzmenko}, Nat Phys \textbf{7}, 48 (2011).

\bibitem{CrasseePRB84_2011}
\bibfnamefont{I.}~\bibnamefont{Crassee},
  \bibfnamefont{J.}~\bibnamefont{Levallois},
  \bibfnamefont{D.}~\bibnamefont{van~der Marel}, \bibfnamefont{A.~L.}
  \bibnamefont{Walter}, \bibfnamefont{T.}~\bibnamefont{Seyller},
  \bibnamefont{and} \bibfnamefont{A.~B.} \bibnamefont{Kuzmenko}, Phys. Rev. B
  \textbf{84}, 035103 (2011).

\bibitem{WuNL12_2012}
\bibfnamefont{Y.~Q.} \bibnamefont{Wu}, \bibfnamefont{K.~A.}
  \bibnamefont{Jenkins}, \bibfnamefont{A.}~\bibnamefont{Valdes-Garcia},
  \bibfnamefont{D.~B.} \bibnamefont{Farmer},
  \bibfnamefont{Y.}~\bibnamefont{Zhu}, \bibfnamefont{A.~A.} \bibnamefont{Bol},
  \bibfnamefont{C.}~\bibnamefont{Dimitrakopoulos}, \bibfnamefont{W.~J.}
  \bibnamefont{Zhu}, \bibfnamefont{F.~M.} \bibnamefont{Xia},
  \bibfnamefont{P.}~\bibnamefont{Avouris}, \bibnamefont{and}
  \bibfnamefont{Y.~M.} \bibnamefont{Lin}, Nano Lett. \textbf{12}, 3062 (2012).

\bibitem{MorimotoPRB86_2012}
\bibfnamefont{T.}~\bibnamefont{Morimoto},
  \bibfnamefont{M.}~\bibnamefont{Koshino}, \bibnamefont{and}
  \bibfnamefont{H.}~\bibnamefont{Aoki}, Phys. Rev. B \textbf{86}, 155426
  (2012).

\bibitem{SadowskiPRL97_2006}
\bibfnamefont{M.~L.} \bibnamefont{Sadowski},
  \bibfnamefont{G.}~\bibnamefont{Martinez},
  \bibfnamefont{M.}~\bibnamefont{Potemski},
  \bibfnamefont{C.}~\bibnamefont{Berger}, \bibnamefont{and}
  \bibfnamefont{W.~A.} \bibnamefont{de~Heer}, Phys. Rev. Lett. \textbf{97},
  266405 (2006).

\bibitem{OrlitaPRL101_2008}
\bibfnamefont{M.}~\bibnamefont{Orlita},
  \bibfnamefont{C.}~\bibnamefont{Faugeras},
  \bibfnamefont{P.}~\bibnamefont{Plochocka},
  \bibfnamefont{P.}~\bibnamefont{Neugebauer},
  \bibfnamefont{G.}~\bibnamefont{Martinez}, \bibfnamefont{D.~K.}
  \bibnamefont{Maude}, \bibfnamefont{A.-L.} \bibnamefont{Barra},
  \bibfnamefont{M.}~\bibnamefont{Sprinkle},
  \bibfnamefont{C.}~\bibnamefont{Berger}, \bibfnamefont{W.~A.}
  \bibnamefont{de~Heer}, \bibnamefont{and}
  \bibfnamefont{M.}~\bibnamefont{Potemski}, Phys. Rev. Lett. \textbf{101},
  267601 (2008).

\bibitem{OrlitaPRL107_2011}
\bibfnamefont{M.}~\bibnamefont{Orlita},
  \bibfnamefont{C.}~\bibnamefont{Faugeras},
  \bibfnamefont{R.}~\bibnamefont{Grill},
  \bibfnamefont{A.}~\bibnamefont{Wysmolek},
  \bibfnamefont{W.}~\bibnamefont{Strupinski},
  \bibfnamefont{C.}~\bibnamefont{Berger}, \bibfnamefont{W.~A.}
  \bibnamefont{de~Heer}, \bibfnamefont{G.}~\bibnamefont{Martinez},
  \bibnamefont{and} \bibfnamefont{M.}~\bibnamefont{Potemski}, Phys. Rev. Lett.
  \textbf{107}, 216603 (2011).

\bibitem{HofmannRSI77_2006}
\bibfnamefont{T.}~\bibnamefont{Hofmann},
  \bibfnamefont{U.}~\bibnamefont{Schade},
  \bibfnamefont{W.}~\bibnamefont{Eberhardt}, \bibfnamefont{C.~M.}
  \bibnamefont{Herzinger}, \bibfnamefont{P.}~\bibnamefont{Esquinazi},
  \bibnamefont{and} \bibfnamefont{M.}~\bibnamefont{Schubert}, Rev. Sci.
  Instrum. \textbf{77}, 63902 (2006).

\bibitem{HofmannTSF519_2011}
\bibfnamefont{T.}~\bibnamefont{Hofmann}, \bibfnamefont{C.~M.}
  \bibnamefont{Herzinger}, \bibfnamefont{J.~L.} \bibnamefont{Tedesco},
  \bibfnamefont{D.~K.} \bibnamefont{Gaskill}, \bibfnamefont{J.~A.}
  \bibnamefont{Woollam}, \bibnamefont{and}
  \bibfnamefont{M.}~\bibnamefont{Schubert}, Thin Solid Films \textbf{519},
  2593– (2011).

\bibitem{SchubertJOSAA20_2003}
\bibfnamefont{M.}~\bibnamefont{Schubert},
  \bibfnamefont{T.}~\bibnamefont{Hofmann}, \bibnamefont{and}
  \bibfnamefont{C.~M.} \bibnamefont{Herzinger}, J. Opt. Soc. Am. A \textbf{20},
  347 (2003).

\bibitem{Hofmannpss205_2008}
\bibfnamefont{T.}~\bibnamefont{Hofmann},
  \bibfnamefont{C.}~\bibnamefont{Herzinger}, \bibnamefont{and}
  \bibfnamefont{M.}~\bibnamefont{Schubert}, phys. stat. sol. (a) \textbf{205},
  779 (2008).

\bibitem{Fujiwara_2007}
\bibfnamefont{H.}~\bibnamefont{Fujiwara}, \emph{Spectroscopic Ellipsometry}
  (John Wiley \& Sons, New York, 2007).

\bibitem{TedescoAPL96_2010}
\bibfnamefont{J.~L.} \bibnamefont{Tedesco}, \bibfnamefont{G.~G.}
  \bibnamefont{Jernigan}, \bibfnamefont{J.~C.} \bibnamefont{Culbertson},
  \bibfnamefont{J.~K.} \bibnamefont{Hite},
  \bibfnamefont{Y.}~\bibnamefont{Yang}, \bibfnamefont{K.~M.}
  \bibnamefont{Daniels}, \bibfnamefont{R.~L.} \bibnamefont{{Myers-Ward}},
  \bibfnamefont{C.~R.} \bibnamefont{Eddy}, \bibfnamefont{J.~A.}
  \bibnamefont{Robinson}, \bibfnamefont{K.~A.} \bibnamefont{Trumbull},
  \bibfnamefont{M.~T.} \bibnamefont{Wetherington}, \bibfnamefont{P.~M.}
  \bibnamefont{Campbell}, \emph{et~al.}, Appl. Phys. Lett. \textbf{96}, 222103
  (2010).

\bibitem{BoosalisAPL101_2012}
\bibfnamefont{A.}~\bibnamefont{Boosalis},
  \bibfnamefont{T.}~\bibnamefont{Hofmann},
  \bibfnamefont{V.}~\bibnamefont{Darakchieva},
  \bibfnamefont{R.}~\bibnamefont{Yakimova}, \bibnamefont{and}
  \bibfnamefont{M.}~\bibnamefont{Schubert}, Appl. Phys. Lett. \textbf{101},
  011912 (2012).

\bibitem{SchubertIRSEBook_2004}
\bibfnamefont{M.}~\bibnamefont{Schubert}, \emph{Infrared Ellipsometry on
  semiconductor layer structures: Phonons, plasmons and polaritons}, vol. 209
  of \emph{Springer Tracts in Modern Physics} (Springer, Berlin, 2004).

\bibitem{TiwaldPRB60_1999}
\bibfnamefont{T.~E.} \bibnamefont{Tiwald}, \bibfnamefont{J.~A.}
  \bibnamefont{Woollam}, \bibfnamefont{S.}~\bibnamefont{Zollner},
  \bibfnamefont{J.}~\bibnamefont{Christiansen}, \bibfnamefont{R.~B.}
  \bibnamefont{Gregory}, \bibfnamefont{T.}~\bibnamefont{Wetteroth},
  \bibfnamefont{S.~R.} \bibnamefont{Wilson}, \bibnamefont{and}
  \bibfnamefont{A.~R.} \bibnamefont{Powell}, Phys. Rev. B \textbf{60}, 11464
  (1999).

\bibitem{SchubertPRB71_2005}
\bibfnamefont{M.}~\bibnamefont{Schubert},
  \bibfnamefont{T.}~\bibnamefont{Hofmann}, \bibnamefont{and}
  \bibfnamefont{J.}~\bibnamefont{{\v S}ik}, Phys. Rev. B \textbf{71}, 35324
  (2005).

\bibitem{LinAPL97_2010}
\bibfnamefont{Y.-M.} \bibnamefont{Lin},
  \bibfnamefont{C.}~\bibnamefont{Dimitrakopoulos}, \bibfnamefont{D.~B.}
  \bibnamefont{Farmer}, \bibfnamefont{S.-J.} \bibnamefont{Han},
  \bibfnamefont{Y.}~\bibnamefont{Wu}, \bibfnamefont{W.}~\bibnamefont{Zhu},
  \bibfnamefont{D.~K.} \bibnamefont{Gaskill}, \bibfnamefont{J.~L.}
  \bibnamefont{Tedesco}, \bibfnamefont{R.~L.} \bibnamefont{Myers-Ward},
  \bibfnamefont{J.}~\bibnamefont{Charles R.~Eddy},
  \bibfnamefont{A.}~\bibnamefont{Grill}, \bibnamefont{and}
  \bibfnamefont{P.}~\bibnamefont{Avouris}, Appl. Phys. Lett. \textbf{97},
  112107 (2010).

\bibitem{SadowskiSSC143_2007}
\bibfnamefont{M.}~\bibnamefont{Sadowski},
  \bibfnamefont{G.}~\bibnamefont{Martinez},
  \bibfnamefont{M.}~\bibnamefont{Potemski},
  \bibfnamefont{C.}~\bibnamefont{Berger}, \bibnamefont{and}
  \bibfnamefont{W.}~\bibnamefont{de~Heer}, Solid State Commun. \textbf{143},
  123  (2007).

\bibitem{OrlitaPRB83_2011}
\bibfnamefont{M.}~\bibnamefont{Orlita},
  \bibfnamefont{C.}~\bibnamefont{Faugeras},
  \bibfnamefont{J.}~\bibnamefont{Borysiuk}, \bibfnamefont{J.~M.}
  \bibnamefont{Baranowski},
  \bibfnamefont{W.}~\bibnamefont{Strupi\ifmmode~\acute{n}\else \'{n}\fi{}ski},
  \bibfnamefont{M.}~\bibnamefont{Sprinkle},
  \bibfnamefont{C.}~\bibnamefont{Berger}, \bibfnamefont{W.~A.}
  \bibnamefont{de~Heer}, \bibfnamefont{D.~M.} \bibnamefont{Basko},
  \bibfnamefont{G.}~\bibnamefont{Martinez}, \bibnamefont{and}
  \bibfnamefont{M.}~\bibnamefont{Potemski}, Phys. Rev. B \textbf{83}, 125302
  (2011).

\bibitem{OrlitaPRL102_2009}
\bibfnamefont{M.}~\bibnamefont{Orlita},
  \bibfnamefont{C.}~\bibnamefont{Faugeras}, \bibfnamefont{J.~M.}
  \bibnamefont{Schneider}, \bibfnamefont{G.}~\bibnamefont{Martinez},
  \bibfnamefont{D.~K.} \bibnamefont{Maude}, \bibnamefont{and}
  \bibfnamefont{M.}~\bibnamefont{Potemski}, Phys. Rev. Lett. \textbf{102},
  166401 (2009).

\bibitem{AbergelPRB75_2007}
\bibfnamefont{D.~S.~L.} \bibnamefont{Abergel} \bibnamefont{and}
  \bibfnamefont{V.~I.} \bibnamefont{Fal'ko}, Phys. Rev. B \textbf{75}, 155430
  (2007).

\bibitem{KoshinoPRB77_2008}
\bibfnamefont{M.}~\bibnamefont{Koshino} \bibnamefont{and}
  \bibfnamefont{T.}~\bibnamefont{Ando}, Phys. Rev. B \textbf{77}, 115313
  (2008).

\end{thebibliography}

\end{document}